\documentclass[arxiv,usenatbib]{iupartex}
\usepackage[T1]{fontenc}
\usepackage{lineno}
\newcommand{\bg}{\begin{linenomath*}\begin{eqnarray}}
\newcommand{\ed}{\end{eqnarray}\end{linenomath*}}
\newcommand{\pao}{
\overline{\partial_e\Omega}
}
\newcommand{\sgn}{
\text{sgn}
}
\newcommand{\textbit}[1]{\textbf{\textit{#1}}}
\newcommand{\inspace}[1]{\vspace*{#1}\noindent}

\title{Incompleteness of Bell's theorem}
\author[J.F.Geurdes]{%
J. F. Geurdes$^{1\cc}$,
\affsep \\
$^1$Independent researcher, pensioner, 2593 NN, 164, Netherlands
\\
$^*$ han.geurdes@gmail.com
}

\corres{J.F.Geurdes}{han.geurdes@gmail.com}

\pubyear{20XX}

\doiheader{XXXXXXX/PAR.20XX.00000}
\date{
	\pSubmit{00.00.0000} 
	\pRevReq{00.00.0000}
	\pLastRevRec{00.00.0000}
	\pAccept{00.00.0000}
	\pPubOnl{00.00.0000}
}
\volume{0}
\volnumber{1}

\begin{document}
\label{firstpage}
\pagerange{\pageref*{firstpage}--\pageref*{lastpage}}
\maketitle

% Abstract of the paper
\begin{abstract}
In the paper it is reported that Bell's correlation formula allows an Einstein local hidden variables explanation.
The key is the application of Petis integration.
\end{abstract}

% Select between one and six entries from the list of approved keywords.
% Don't make up new ones.

%%%%%%%%%%%%%%%%%%%%%%%%%%%%%%%%%%%%%%%%%%%%%%%%%%

%%%%%%%%%%%%%%%%% BODY OF PAPER %%%%%%%%%%%%%%%%%%

\section{Introduction}
The foundational discussion started by Einstein about the existence of extra parameters that would restore locality and causality to quantum mechanics \cite{EPR} and a version more to Einstein's taste \cite{Howard}, appears to be decided by experiment \cite{Aspect}. The outcome did not supported Einstein's views.
Nevertheless, the present author \cite{Geurdes} disagrees with Aspect's statistics in the experiment.
The reason is as follows.
Given $x$ in $ [0,2 \pi)$ is the angle between Alice's $\mathbf{a}$ unit length parameter vector and Bob's unit length parameter vector $\mathbf{b}$. 
Given the angle is measured e.g. from $\mathbf{a}$ to $\mathbf{b}$.
Aspect then requires: 
\bg\label{noasp}
\begin{array}{ll}
1. & \cos(x)= P(x,=) - P(x,\neq)\\
1a. & P(x,=)=N(x,=)/N\\
1b. & ~P(x,\neq)= N(x,\neq)/N\\
1c. & N(x,=)=N(x,+,+)+N(x,-,-)\\
1d.& N(x,\neq)=N(x,+,-)+N(x,-,+)\\
1e. & N=N(x,=)+N(x,\neq)\\
2.& \cos(x)=1-2\sin^2(x/2), x \in [0,2 \pi)\\
3. & P(x,=)+P(x,\neq) =1\\
4. & P(x,\neq)=\sin^2(x/2), x \in [0,2 \pi)
\end{array}
\ed
The $+$ and $-$ denote the  measured polarization. I.e. $(+,+)$ denotes that both Alice and Bob measured an equal polarization and the polarization is denoted with $+$, etc. 
The first equation above asks if measurements of $=$ and $\neq$ polarization of entangled photon pairs can produce the quantum correlation.
Note that the law of large numbers is applied to avoid the definition of a certain model of local extra parameters in the experiment.
The left-hand in 4. is a data probability (estimate) determined by Nature. But note please, the right-hand isn't a probability function. It isn't monotone non-descending for $x \in [0,2\pi)$. 
\\\\
\noindent
Another noreworthy study is \cite{Weihs}.
In Aspect's study and Weihs's, the CHSH inequality  \cite{CHSH} is employed. 
It can be  derived from Bell's correlation formula \cite{Bell}.
The CHSH inequality is a means to test the outcome of the experiment.
However, the requirement in (\ref{noasp}) shows that this requirement can not be met in the data.
It requires violations of the Kolmogorov axioms.
Aspect's experiment proved nothung.
It, in turn, implies that theoretical extensions of the Bell correlation formula might still be viable to produce an Einstein extra parameter explanation of entanglement.
Bell's correlation formula is intended to be a general function describing all possible Einstein local hidden variables models.
Then if we look at the equation we might obviously also have the following format for Bell's correlation
\bg\label{b1}
P(a,b)=
\int d\lambda 
\int d\lambda' 
\rho_{\mathbf{a}}(\lambda)
\rho_{\mathbf{b}}(\lambda')
A(\lambda)B(\lambda')
\ed
In this particular example, the setting parameter vector $\mathbf{a}$ determines the density of the hidden variables $\lambda$ at measurement instrument $A$, represented by measurement function $A(\lambda)$.
Similar configuration goes for Bob.
The measurement functions do not depend on the settings.
The density of the hypothetical hidden variables are determined by the settings.
\\\\
\noindent
From this we can see that locality can be maintained. Setting vectors determine the density while the entangled particles carry the $(\lambda, \lambda')$.
It is the generality of Bell's formula plus the locality preserving manner of probability density that will be researched further. 
If $\mathbf{a}$ influences the $\rho_{\mathbf{a}}(\lambda)$ but not the $\lambda'$, locality seems to be warranted. 
But do note that this form of Bell's correlation formula is not forbidden \emph{and} that, in generality, it disallows a CHSH type of inequality, viz. \cite{CHSH}.
Same story for $\mathbf{b}$ influencing $\rho_{\mathbf{b}}(\lambda')$ but not $\lambda$.  
\\\\
\noindent
The author acknowledges that many studies have been performed on Bell's theorem. 
The author is sorry for not referring to them.
But, frankly speaking, they appear to be unnessecary when looking at the trust in the Aspect experiment they have generated.
Aspect's experiment is statistically flawed and no Nobel prize will undo this statistical - probabilistic fact.
The present paper is absolutely not a physics history paper but an attempt to overthrow the hegemony of a theorem that is incomplete.
\section{Petis integral and the Krein-Milman theorem from measure theory}
In \cite{Masani} the Krein-Milman theorem is presented in an architectonic setting of mathematical structure. 
Here, we are only interested in the application of that theorem to Bell's correlation case.
\\\\
\noindent
On page 116 of the work of  \cite{Masani} we can read that a topological local convex vector space $X$ is necessary.
In our case we may take  $X\subset\mathbb{R}^3$.
The metric is Euclidian.
A convex space is a space where each point can be connected with another point via a straight line.
This is possible in $X\subset \mathbb{R}^3$.
Subsequently, we are in need of a compact convex subspace, $\Omega$, with, $\Omega\subset X$..
This space is defined by
\bg\label{b2}
\Omega=
\left\{ 
\mathbf{x}\in\mathbb{R}^3
~
|
~
x_1^2+x_2^2+x_3^2\leq 1
\right\} 
\ed
We have, $\mathbf{x}^T\cdot\mathbf{x} = ||\mathbf{x}||= x_1^2+x_2^2+x_3^2$.
The space $\Omega$ is compact because it is closed and bounded.
It is a convex subset of $\mathbb{R}^3$.
The boundary $\partial_e\Omega$  of $\Omega$ is given by
\bg\label{b3}
\partial_e\Omega = 
\left\{ 
\mathbf{x}\in\mathbb{R}^3
~
|
~
x_1^2+x_2^2+x_3^2\ = 1
\right\} 
\ed
Note that $\partial_e\Omega \subset \Omega$.
Hence, all $\omega_0 \in \partial_e\Omega$ are obviously  also, $\omega_0 \in\Omega$.
The closure of $\partial_e\Omega$ is $\pao=\partial_e\Omega$. 
According to the Krein-Milman theorem there is a (probability) measure possible such that for arbitrary $\omega_0 \in\Omega$
\bg\label{km1}
\mu_{\omega_0 }(\pao)=1,
\\\nonumber\\\nonumber
\omega_0=
\int_{\pao} \omega \mu_{\omega_0}(d\omega)
\ed
The integral in the above equation is a Petis integral viz. \cite{Masani} based on a probability measure $\mu_{\omega_0}(d\omega)$.
Vector measures are discussed in a textbook \cite{VectMeas}.
Some simpler formulations can be found in \cite{Basu}.
Because $\partial_e\Omega \subset \Omega$, we may assume that there is a measure $\mu_{\omega_0}$ such that all, $\omega_0 \in \pao\subset \Omega$ are covered.
The role of $\omega_0$
is similar to e.g. the mean in a normal Gaussian density.
Then, it is easy to understand that a seting vector $\mathbf{a}$ is recoveredd with $\mu_a$ and that $\mu_a(\pao)=\int_{\pao} \mu_{\omega_0}(d\omega)=1$, for arbitrary $\omega_0 \in \pao \subset \Omega$.
And should one insist that $\omega_0$ is from $\Omega - \pao$, then an arbitrary close to $\omega_0 \in \pao$, e.g. $\omega_1 \in \Omega - \pao $ applies.
The classical result of Minkovski \cite{Phelps} is:
\\\\\noindent
\emph{
If x is an element of the compact convex set $\Omega$, then x is a finite combination of points from the extreme $x_1, x_2,\dots x_n$, such that for positive numbers $\mu_1, \mu_2,\dots \mu_n$ with $\sum_{k=1}^n \mu_k =1$, we have $x=\sum_{k=1}^n \mu_k x_k$
}
\\\\\noindent
In the following discussion it is easier to allow vectors from  $\pao$ even though a infinitely small approach of it might be more accurate..
To be absolutely clear, with respect to (\ref{b2}) and (\ref{b3}),  $\pao \subset \Omega$. 

\subsection{The measurement functions}
Let's for a moment return to (\ref{b1}) and acknowledge that the measurement functions $A$ and $B$ can depend on $\omega$ and $\omega'$. 
In the physics of this situation one must not forget that the two particles arise from a single source and that one of the particles is allowed to carry "something" with it referring to the other particle.
This concept does not violate the principle of locality. In (\ref{b1}) the measurement functions do not depend on the parameter vectors $\mathbf{a}$ and $\mathbf{b}$. 
And the densities, reflected in the Petis measures of (\ref{km1}), i.e.  the $\mu_{\mathbf{a}}$ and $\mu_{\mathbf{b}}$, are local.
\\\\
\noindent
Let us also note that the averages of $A$ and $B$ are expected to be zero and the variances of $A$ and $B$ are expected to be unity.
This necessitates to introduce two new extra parameters.
The first one, $x$ has a Gaussian normal density.
I.e.
\bg\label{ex1}
\frac{1}{\sqrt2\pi} \int_{-\infty}^\infty e^{-x^2/2} dx =1,
\\\nonumber
\frac{1}{\sqrt2\pi} \int_{-\infty}^\infty e^{-x^2/2} \sgn(x) dx =0,
\\\nonumber
\frac{1}{\sqrt2\pi} \int_{-\infty}^\infty e^{-x^2/2} \sgn^2(x) dx =1
\ed
The $\sgn^2(x)=1$ for $x\neq 0$ 
and one point taken out of the integral doesn't change its value.
Secondly, we introduce the variable $y$ with, $-1\leq \xi \leq 1$, such that
\bg\label{y}
\frac{1}{2}\int_{-1}^1 \sgn \{\xi - y \} dy =
\frac{\xi}{2} +\frac{1}{2}-\frac{1}{2} + \frac{\xi}{2}=
\xi
\ed
And so we are then able to define the two measurement functions $A$ and $B$ as they occur in (\ref{b1})
\bg\label{ab}
A=\sgn(x)
\sgn(|\omega^T\cdot\omega'|^{1/2} - y)
\sgn(\sgn(\omega^T\cdot\omega') - \eta)
\\\nonumber
B=\sgn(x)
\sgn(|\omega^T\cdot\omega'|^{1/2} - z)
\ed
The inner product, $(\omega^T\cdot\omega')=\sum_{k=1}^3 \omega_k \omega'_k$. 
It has $-1\leq (\omega^T\cdot\omega')\leq 1$ for the inner product of every two points on $\pao$.
\subsection{The integration}
Bell's correlation arises from an integration.
And so we define, together with the Petis integrals, for the parameters the following 
expectation (integration) value when e.g. $\mathbf{a}$ and $\mathbf{b}$ are employed. 
\bg\label{in1}
E(f)
=
\int_{\pao} \mu_{\mathbf{a}}(d\omega)
\int_{\pao} \mu_{\mathbf{b}}(d\omega')
\int_{-1}^1\frac{dy}{2}
\int_{-1}^1\frac{d\eta}{2}
\int_{-1}^1\frac{dz}{2}
\int_{-\infty}^\infty \frac{dx}{\sqrt{2\pi}} e^{-x^2/2} 
f(\omega,\omega',y,\eta,z,x)
\ed
Then with the definition in (\ref{ab}) and $\int_{\pao} \mu_{\mathbf{a}}(d\omega)=\mu_{\mathbf{a}}(\pao) = 1$, etc,
we find, with (\ref{in1}),  $E(1)=1$ and 
\bg\label{exp}
E(A)=E(B)=0
\\\nonumber
E(A^2)=E(B^2)=1
\ed
for every $\mathbf{a}$ and $\mathbf{b}$.
If we then note that
$-1\leq (\omega^T\cdot\omega')\leq 1$
and (\ref{y}) is valid, while (\ref{in1}) also applies, it follows,
\bg\label{eab}
E(AB)=
\int_{\pao} \mu_{\mathbf{a}}(d\omega)
\int_{\pao} \mu_{\mathbf{b}}(d\omega')
\sgn(\omega^T\cdot\omega')
|\omega^T\cdot\omega'|^{1/2}
|\omega^T\cdot\omega'|^{1/2}
=
\\\nonumber
=
\int_{\pao} \mu_{\mathbf{a}}(d\omega)
\int_{\pao} \mu_{\mathbf{b}}(d\omega')
\sgn^2(\omega^T\cdot\omega')
(\omega^T\cdot\omega')
=
\\\nonumber
=
\left(
\int_{\pao} \mu_{\mathbf{a}}(d\omega) \omega^T
\right)
\cdot
\left(
\int_{\pao} \mu_{\mathbf{b}}(d\omega') \omega'
\right)
=
\\\nonumber
=
\left(
\int_{\pao} \mu_{\mathbf{a}}(d\omega) \omega
\right)^T
\cdot
\left(
\int_{\pao} \mu_{\mathbf{b}}(d\omega') \omega'
\right)=
\mathbf{a}^T\cdot \mathbf{b}
\ed
The upper $T$ index in (\ref{eab}) is to stress the transpose of the vector in the inner product.
In addition, $|p|=p\sgn(p)$, and $p\in\mathbb{R}$.
\section{Conclusion \& discussion}
The result of (\ref{eab}) is the quantum correlation.
Contrary to common folklore, it can be derived from a form of Bell correlation.
Generally speaking, no CHSH inequality can be obtained from the form in (\ref{in1}).
The quantum correlation, $\mathbf{a}^T\cdot \mathbf{b}$,  can be based on local parameters.
Note the fact that $\omega'$ is carried towards $A$ but relates to the $B$ parameter $\mathbf{b}$.
And vice versa.
This can be explained by the fact that the particle moving towards $A$ can be carrying $\omega'$ towards $A$.
This is not a violation of locality because the particles to $A$ and $B$ had contact in the source $S$.
After the particles separate and  leave the source, e.g. the particle towards $A$ may carry "information" of the particle towards $B$.  
It is, therefore, not a non-local hidden parameter explanation. 
This is supported by the following. 
\\\\
\noindent
When the measurements on $A$ and $B$ are performed, the $E(AB)$ becomes macroscopic and subsequently simultaneously knowable to both Alice and Bob. 
One might even introduce a Wigner's friend type of a person $W$ who knows the $\mathbf{a}$ and $\mathbf{b}$ before the two entangled particles go separate ways but does not convey them to the outside world..
If we suppose that one entangled pair can be measured (it can not in practice) then after the measurements, the $\mathbf{a}$ and $\mathbf{b}$ are known.
Or perhaps sharper, $W$ is doing the computation of the correlation with the Bell correlation formula presented in the above, e.g., (\ref{eab}).
\\\\\noindent
The fact that in practice it is not possible to measure one single pair is not a matter of principle.
Therefore, the outcome $E(AB)=\mathbf{a}^T\cdot \mathbf{b}$ becomes known.
The notion that the experiment refuted the existence of this kind of models is, in its turn, refuted by the facts presented in the series of equations in (\ref{noasp}).
Never any serious opposition was given to what is presented in the equations of (\ref{noasp}).
%%%Ruses don't deal with statistical errors in key experiments.
\\\\
\noindent
It must be noted, however,  that a many particle measurements objection is not a principle point.
After a one pair two-particle measurement, observers compare the result for $\mathbf{a}$ and $\mathbf{b}$ and find $\mathbf{a}^T\cdot \mathbf{b}$ from their Bell correlation formula in equation (\ref{eab}).
This surely supports an Einsteinian view of entanglement. 
The classical result of Minkovski, viz. \cite{Phelps} is a genuine mathematical possibility to get there.
\section*{Declarations}
No funding. No conflict of interest. No clinical application. No animals empolyed in this research. No human data in this research.

%%%%%%%%%
%%%%%%%%%%%%%%%%%%%% REFERENCES %%%%%%%%%%%%%%%%%%
% The best way to enter references is to use BibTeX:

%\bibliographystyle{mnras}
%\bibliography{example} % if your bibtex file is called example.bib

%\bibliographystyle{ieeer}

\end{document}